\documentclass[secnumarabic,onecolumn,showkeys,amsmath,amssymb]{revtex4} 

\usepackage{amsmath}    	
\usepackage{amssymb}
\usepackage{graphicx}   		

\usepackage{color}      		
\usepackage{subfigure}

 \pdfoutput=1

\begin{document}

\title{A Method to Calculate the Exit Time in Stochastic Simulations}
\author{Basil S. Bayati}
\affiliation{Institute for Disease Modeling, Intellectual Ventures, 3150 139$^{\text{th}}$ Ave SE, Bellevue, WA 98004, USA }

\date{\today}

\begin{abstract}
A novel method is presented to compute the exit time for the stochastic simulation algorithm.  The method is based on the addition of a series of random variables and is derived using the convolution theorem.  The final distribution is derived and approximated in the frequency domain.  The distribution for the final time is transformed back to the real domain and can be  sampled from in a simulation.  The result is an approximation of the classical stochastic simulation algorithm that requires fewer random variates.  An analysis of the error and speedup compared to the stochastic simulation algorithm is presented.   
\end{abstract}


\maketitle
 
\section{Introduction}
Stochastic processes are intrinsic to complex physical phenomena that range from stellar dynamics  \cite{Chandrasekhar:1943} to epidemiology \cite{Dietz:1967}.  An important example is stochastic chemical kinetics which describes the time evolution of chemically reacting systems by taking into account the fact that molecules are discrete entities that exhibit randomness in their dynamical behavior. The transition probabilities of such processes obey the Chapman-Kolmogorov equation, which in turn is equivalent to the master equation \cite{Kampen:2007,Gardiner:2009}. The number of variables in the master equation is large for all but the simplest systems, so analytical or direct numerical integration methods are usually impractical. Alternatively, Monte Carlo samples of the stochastic process can be numerically generated, via stochastic simulation algorithms (SSAs) \cite{Gillespie:1976,Gillespie:1977}, so that the only error introduced is the sampling error.

There are a variety of simulation \cite{Gillespie:2009} and approximation \cite{Kampen:2007,Gardiner:2009} methods available, and the appropriate method will depend on the system and the specific question posed.  At one end of the spectrum is the stochastic simulation algorithm \cite{Gillespie:1976,Gillespie:1977}, which is a method that produces exact random variates from the master equation such that the probability density function can be reconstructed with an error of $\mathcal{O}(N^{-1/2})$ due to only the sampling error, where $N$ is the number of random variates.  Additionally, there are leaping methods \cite{Gillespie:2001,Auger:2006} that accelerate the running time of the stochastic simulation algorithm while also accruing an additional error that is a function of the time-step \cite{Cao:2006} or the number of discrete quantities in the system \cite{Anderson:2011}.  At the other end of the spectrum is the chemical Langevin equation, which is a random variate from the Fokker-Planck equation, and the continuum reaction rate equations that dispense with fluctuations altogether. 

In the classical stochastic simulation algorithm \cite{Gillespie:1976,Gillespie:1977}, two uniformly distributed random numbers are required per time-step: 1) to select which reaction occurs and 2) to select the time-step.  Here we present a method to reduce the number of random variates needed to compute the time-steps.  The method entails simulating the reactions \emph{without time} and computing the final time once a desired state has been reached.  For example, in chemical kinetics, the desired state might be a high concentration of a product.  The distribution of the final time is derived here, as well as an approximation that is useful for simulations.  It is shown that the final time of the simulation can be computed using a much smaller number of random variates.  

In section 2 we first review the classical stochastic simulation algorithm, then derive the method.  In section 3 we provide results for a simple numerical example and in section 4 we provide concluding remarks on this work.  An appendix section is provided that includes the details of the mathematics used in section 2.

\section{Derivation \& Method}
\subsection{Stochastic Simulation Algorithm}
Chemical reactions can be written in the form 

\begin{equation}
\label{eq:Reaction}
	r_1 X_1 + r_2 X_2 + \ldots \xrightarrow{k} g_1 X_1 + g_2 X_2 + \ldots 
\end{equation}
where $k$ is the reaction rate, $r_k$ represents the number of $X_k$ molecules that participate in the reaction, and $g_k$ the number of $X_k$ products.    Let $\mathbf{X}(t) = (X_1(t), X_2(t), \ldots)^T$ be a realization of the stochastic process, where $\mathbf{X}(t) \in S$, where $S$ denotes an enumeration of every possible state.  The \emph{stoichiometric vector} for reaction \eqref{eq:Reaction} is $\boldsymbol{\nu}_i = (g_1 - r_1, g_2 - r_2, \ldots)^T$ so that if the current state is $\mathbf{X}(t)$ and reaction $i$ occurred within $\text{d}t$, then $\mathbf{X}(t+\text{d}t) = \mathbf{X}(t)+\boldsymbol{\nu}_i$.    

The sample space $S$ of this stochastic process can be visualized as an integer lattice $\mathbb{Z}^d$, where the dimension  $d$ is the number of species in the system. Usually the sample space is smaller than the whole of $\mathbb{Z}^d$, thus it is $[0, \Omega]^d$, where $\Omega$ is the total number of particles in the system.  A propensity is defined for a reaction indexed by $i$ as 
\begin{equation} 
\label{eq:Propensities}
a_{i}(\mathbf{X}(t)) \triangleq  k \Omega  \prod_{j} \Bigg\lbrace \frac{ \mathbf{X}_j(t) \left( \mathbf{X}_j(t)-1)\cdots (\mathbf{X}_j(t)-r_j+1 \right)}{ \Omega^{r_j}} \Bigg\rbrace,
\end{equation}
where $a_i(\mathbf{X}(t))~\text{d}t$ is the probability of a reaction.  Intuitively, the product appears in \eqref{eq:Propensities} since the collisions of molecules is assumed to be independent, and the factor of $\Omega^{r_j}$ is needed so that $a_i(\mathbf{X}(t))$ has units of $[time]^{-1}$.  

The Stochastic Simulation Algorithm (SSA) is a Monte Carlo method for the simulation of chemical reactions.  SSAs deal with a realization of the time-dependent stochastic processes, namely a trajectory $\mathbf{X}(t) \in S$.  The process is simulated over time by the following update scheme: 
\begin{eqnarray}
\mathbf{X}(t+\tau) &=& \mathbf{X}(t) + \boldsymbol{\nu}_{k}.
\end{eqnarray}
 In the classical formulation of the Stochastic Simulation Algorithm (SSA) \cite{Gillespie:1976,Gillespie:1977}, the probability of a reaction $k$ with a time-step $\tau$ is chosen from the joint probability density function
\begin{equation}
\label{eq:ssaT}
p(\tau, k) = a_k(t) e^{-a_{0}(t) \tau},
\end{equation}
where the propensities are defined by equation \eqref{eq:Propensities} and the total propensity is
\begin{equation}
\label{eq:a0}
a_{0}(t) \triangleq \sum_k a_{k} (\mathbf{X}(t)).  
\end{equation}
Equation \eqref{eq:ssaT} can be decomposed as  $p(\tau, k)$ $=$ $p(\tau) p(k)$, where
\begin{eqnarray}
\label{eq:ssa1}
p(\tau) &=& a_0(t) e^{-a_{0}(t) \tau}, \\ 
\label{eq:ssa2}
p(k) &=& \frac{a_k (t)}{a_{0}(t)},
\end{eqnarray}
which amount to calculating the time-step in which a reaction occurred and finding the index of the reaction that occurred within said time-step.  The inverse transform sampling method \cite{Devroye:1986} is used to sample $\tau$ and $k$ from equations \eqref{eq:ssa1} and \eqref{eq:ssa2}.  For instance, 
\begin{equation}
	r_1 = \int_{0}^{\tau} p(\tau')~\text{d}\tau', 
\end{equation}   
where $r_1$ is sampled from a uniform distribution in the range $[0,1)$.  Solving for $\tau$ yields
 \begin{equation}
 \label{eq:SampleTau}
	\tau = -\frac{1}{a_0(t)}~\text{ln}(r_1).
\end{equation}    
The value for $j$ is the integer for which 
 \begin{equation}
 \label{eq:SampleJ}
	\sum_{\beta=1}^{j-1} a_{\beta}(t) < r_2 a_0(t) \leq \sum_{\beta =1}^{j} a_{\beta}(t). 
\end{equation}    
where $r_2$ is another sample from a uniform distribution in the range $[0,1)$.

The algorithm is: 
 $0$. Initialize the time $t=0$ and the system's state $\mathbf{X}=\mathbf{X}_0$.
\begin{enumerate}
\item With the system in state $\mathbf{X}$ at time $t$, evaluate all of the propensities $a_j(t)$ (equation \eqref{eq:Propensities}) and their sum $a_0(t)$ (equation \eqref{eq:a0}).
\item Generate values for $\tau$ and $j$ where $\tau$ is an exponential random variable with parameter $a_0(t)$ (equation \eqref{eq:SampleTau}) and $j$ is a discrete random variable with $P(j=k)=\frac{a_k(t)}{a_0(t)}$ (equation \eqref{eq:SampleJ}).
\item Execute the next reaction by replacing $t\rightarrow t+\tau$ and $\mathbf{X}\rightarrow \mathbf{X}+\boldsymbol{\nu}_{j}$ where $\boldsymbol{\nu}_{j}$ is the stoichiometric vector that denotes the change induced by reaction $j$.
\item Record $(\mathbf{X},t)$ as desired. Return to step 1, or else end the simulation.
\end{enumerate}

\subsection{Exit Time Method}

Consider the transition from an initial state $\boldsymbol{X}_1$ to the boundary $\boldsymbol{X}_{n+1}$, where $\boldsymbol{X}_{n+1}$ could denote a high concentration of a particular product.  The joint probability is 
\begin{equation}
p(t, \boldsymbol{X}_{n+1}) = p(t | \boldsymbol{X}_1 \rightarrow  \cdots \rightarrow  \boldsymbol{X}_{n+1} ) p( \boldsymbol{X}_1 \rightarrow  \cdots \rightarrow  \boldsymbol{X}_{n+1}).  
\end{equation} 
Since $p(t | \boldsymbol{X}_1 \rightarrow  \cdots \rightarrow  \boldsymbol{X}_{n+1} )$ depends on the total propensity of the states at each iteration, we will write this as $p\left(t | \lambda_1, \ldots, \lambda_n  \right)$, where $\lambda_i = a_0(\boldsymbol{X}_i,t_i)$.  
Below we will derive an expression for $p\left(t | \lambda_1, \ldots, \lambda_n  \right)$, which is known as the \emph{exit time} or \emph{hitting time} of a stochastic simulation.

\begin{figure}[htp]
  \begin{center}
     \includegraphics[width= 1\textwidth] {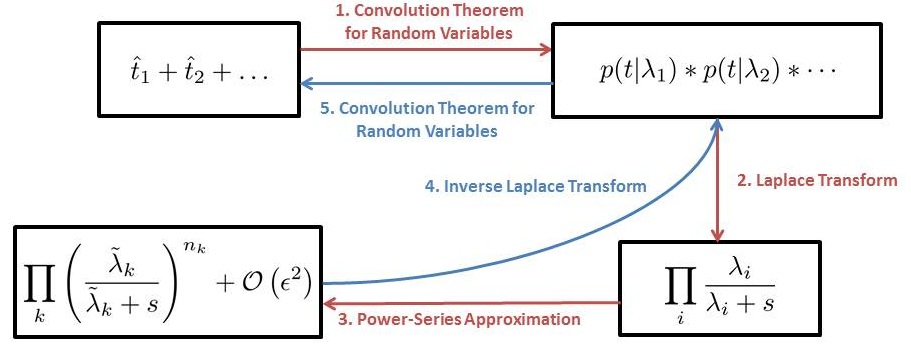}
     	\caption[]{ Schematic of the derivation.  The numbers denote the steps.  The derivation consists of using the convolution theorem for the addition of random variables, transforming to the frequency domain, making an approximation, then transforming back as to obtain a modified and smaller set of random variables. }
	 \label{fig:schematic} 
  \end{center}
 \end{figure}

The derivation will follow the schematic shown in Figure \ref{fig:schematic}

At each iteration in the stochastic simulation algorithm, we will increment time by sampling from an exponential random variable with a density of $p_i(t; \lambda_i)$.  Therefore, we are interested in finding the sum of $n$ exponential random variables that are sampled from $p_i(t; \lambda_i)$ for $i = 1, \ldots, n$, where $\lambda_i = a_0(\boldsymbol{X}_i,t_i)$.  We begin by defining the exponential distribution: 
\begin{eqnarray}
p_i(t; \lambda_i) \triangleq \lambda e^{-\lambda_i t},   
\end{eqnarray}
where $t \in [0, \infty)$ and $\lambda_i > 0$.  

\subsubsection*{Step 1: Convolution Theorem.}
We will find the sum of the variables by using the convolution theorem.  Let $\hat{t}$ be the sampled time after $n$ iterations.  In the stochastic simulation algorithm we have
\begin{eqnarray}
\hat{t} = \mathcal{E}\left(\frac{1}{\lambda_1} \right) + \cdots + \mathcal{E}\left(\frac{1}{\lambda_n} \right) + \mathcal{O}\left(N^{-1/2} \right), 
\end{eqnarray}
where $N$ is the number of Monte Carlo samples and $\mathcal{E}\left(\frac{1}{\lambda_i} \right)$ is a random variate sampled from $p_i(t; \lambda_i) = \lambda e^{-\lambda_i t}$.  Let $*$ denote a convolution, then the convolution theorem \cite{Springer:1979} states that
\begin{eqnarray}
\label{eq:conv}
p\left(t | \lambda_1, \ldots, \lambda_n  \right)&=&  p_1(t; \lambda_1) * \cdots * p_n(t; \lambda_n) 
\end{eqnarray}
for the final distribution for the random variate $\hat{t}$, i.e. $\hat{t} \sim p_1(t; \lambda_1) * \cdots * p_n(t; \lambda_n)$.  
  
\subsubsection*{Step 2: Laplace Transform.}
We begin by transforming equation \eqref{eq:conv}
\begin{eqnarray}
 \mathcal{L}\left\{  p_1(t; \lambda_1) *   \ldots  * p_n(t; \lambda_n)  \right\}(s) &=& \prod_{i=1}^n \mathcal{L}\left\{  p_i(t; \lambda_i)  \right\}(s), 
\end{eqnarray}
where we have used Laplace's transform in lieu of Fourier's since $t \in [0, \infty)$ and we have used the convolution property of the transform.  We find that 
\begin{eqnarray}
\mathcal{L} \left\{  p_i(t; \lambda_i)  \right\}(s) &\triangleq& \int_0^{\infty} p_i(t; \lambda_i) e^{-s t}~\text{d}t \\
&=& \frac{\lambda_i}{\lambda_i + s},
\end{eqnarray}  
where $s \in \mathbb{C}$.  Therefore, we can write 
\begin{eqnarray}
 \mathcal{L}\left\{  p_1(t; \lambda_1) *   \ldots  * p_n(t; \lambda_n)  \right\}(s) &=&  \prod_{i=1}^n \mathcal{L}\left\{  p_i(t; \lambda_i)  \right\}(s)   \\
&=&  \prod_{i=1}^n  \frac{\lambda_i}{\lambda_i + s}.  
\end{eqnarray}
The analytical expression can be found in two cases: 1) if $\lambda_i \neq \lambda_j~\forall i,j$, then (see Appendix section \ref{subusb:DerivationOfHypoexponentialDistribution})
\begin{eqnarray} \label{eq:hyp}
p\left(t| \lambda_1, \ldots, \lambda_n \right) =  \mathcal{L}^{-1}\left\{   \prod_{i=1}^n  \frac{\lambda_i}{\lambda_i + s}   \right\}  =  \sum_{i=1}^n  \left(  \prod_{\substack{j=1\\j \neq i}}^n \frac{\lambda_j}{\lambda_j - \lambda_i} \right)   \lambda_i e^{-\lambda_i t}, 
\end{eqnarray}
and 2) if  $\lambda = \lambda_i~\forall i$, then (see Appendix section \ref{sec:GammaFrequency})
\begin{eqnarray} \label{eq:gammaEq}
p\left(t| \lambda_1, \ldots, \lambda_n \right) = \mathcal{L}^{-1}\left\{   \prod_{i=1}^n  \frac{\lambda}{\lambda + s}   \right\}  = 
\frac{\lambda^n t^{n-1} e^{-\lambda t}}{\Gamma(n)}, 
\end{eqnarray}
which is the Erlang (a.k.a. Gamma) distribution.  In general the residue theorem (see Appendix section \ref{sec:residueTheory}) could be used to find the inverse Laplace transform, but symbolic differentiation would be necessary so this is avoided.  Since the total propensity will change over time, we are interested in drawing random variates from equation \eqref{eq:hyp}, and we therefore need to find the inverse function.  However, equation \eqref{eq:hyp} has no inverse (see Appendix section \ref{sec:RandomVariatesFromPDF}) and, moreover, it is numerically unstable rendering it impractical.  

\subsubsection*{Step 3: Approximation.}
We therefore want to approximate $ \prod_{i=1}^n  \frac{\lambda_i}{\lambda_i + s}$ so as to obtain the Erlang distribution, i.e. equation \eqref{eq:gammaEq}.  To illustrate the approximation, we let  $|\lambda_1 - \lambda_2| = 2 \epsilon$ where $\tilde{\lambda}$ is chosen such that $\lambda_1 = \tilde{\lambda}+\epsilon$ and $\lambda_2 = \tilde{\lambda}-\epsilon$, then (see Appendix section \ref{sec:Approx})
\begin{eqnarray}
\left(\frac{\lambda_1}{\lambda_1 + s}\right) \left(\frac{\lambda_2}{\lambda_2 + s}\right) = \left( \frac{\tilde{\lambda}}{\tilde{\lambda} + s} \right)^2 + \mathcal{O}\left( \epsilon^2 \right), 
\end{eqnarray}
where $ \mathcal{O}\left( \epsilon^2 \right)$ denotes an error on the order of $\epsilon^2$.
By grouping together $\lambda_i$s that differ by $\epsilon$ into disjoint sets, we can write 
\begin{eqnarray}
\prod_{i = 1}^{n} \left(\frac{\lambda_i}{\lambda_i + s}\right) = \prod_{k=1}^{m} \left(\frac{\tilde{\lambda}_k}{\tilde{\lambda}_k + s}\right)^{n_k} +\mathcal{O}\left( \epsilon^2 \right), 
\end{eqnarray}
where $m \ll n$, and each $\tilde{\lambda}_k$ and $n_k$ are chosen according to $\epsilon$ (see Appendix section \ref{sec:Epsilon}).  Therefore, 
\begin{eqnarray}
\mathcal{L}\left\{ p\left(t| \lambda_1, \ldots, \lambda_n \right) \right\}(s) =  \prod_{k=1}^{m} \left(\frac{\tilde{\lambda}_k}{\tilde{\lambda}_k + s}\right)^{n_k} +\mathcal{O}\left( \epsilon^2 \right).  
\end{eqnarray}

\subsubsection*{Step 4: Inverse Laplace Transform.}
Transforming back, we have
\begin{eqnarray} 
p\left(t | \lambda_1, \ldots, \lambda_n  \right) &=& \mathcal{L}^{-1}\left\{\prod_{k=1}^{m} \left(\frac{\tilde{\lambda}_k}{\tilde{\lambda}_k + s}\right)^{n_k} +\mathcal{O}\left( \epsilon^2 \right) \right\}(t) \\
&=& \Gamma \left(\frac{1}{\tilde{\lambda}_1}, n_1 \right) * \cdots *  \Gamma \left(\frac{1}{\tilde{\lambda}_m}, n_m \right) +\mathcal{O}\left( \epsilon^2 \right).  
\end{eqnarray}

\subsubsection*{Step 5: Convolution Theorem.}
We want to generate a sample $\hat{t} \sim p\left(t | \lambda_1, \ldots, \lambda_n  \right)$ for each trajectory, therefore we sample from the sum of the distributions:  
 \begin{eqnarray}
\label{eq:finalSample}
\hat{t}= \sum_{k=1}^{m} \gamma \left(\frac{1}{\tilde{\lambda}_k}, n_k \right) +\mathcal{O}\left( \epsilon^2 \right) + \mathcal{O}\left( N^{-1/2} \right), 
\end{eqnarray}
where $\gamma \left(\frac{1}{\tilde{\lambda}_k}, n_k \right) \sim  \Gamma \left(\frac{1}{\tilde{\lambda}_k}, n_k \right)$ is a random variate from a Gamma distribution with a scale parameter of $\frac{1}{\tilde{\lambda}_k}$ and a shape parameter of $n_k$, and $N$ is the number of Monte Carlo samples.    We can now simulate a chemical system without time until a desired state has been reached, and use equation \eqref{eq:finalSample} to compute the final time of the simulation once we have grouped together the propensities as shown in the Appendix section \ref{sec:Epsilon}.  

\section{Results}
\label{sec:Results}

Here we consider the most elementary of nonlinear systems which has been used to model many disparate physical processes ranging from nuclear reactions \cite{Canosa:1969} to epidemics \cite{Dietz:1967}:
\begin{eqnarray}
\label{eq:StoI}
S+I &\xrightarrow{\beta}& 2 I,  \\
I &\xrightarrow{\gamma}& R.
\end{eqnarray}
 In epidemics, this is the canonical SIR model \cite{Kermack:1927}, upon which more detailed models that include age- and spatially-dependent processes are built.  The reproductive number is defined $R_0 \triangleq \beta / \gamma$, and $S$, $I$, and $R$ denote the susceptible, infectious, and recovered persons, respectively.  This process models the event in which a  susceptible person comes into contact with a infectious person at a rate $\beta$ and results in two infectious persons.  We use $\beta=3/2, \gamma=1$ yielding $R_0=3/2$, $\Omega=100$ and set the exit condition as $R=85$, i.e. the exit vector is $\mathbf{X}_{\text{exit}}=(.,.,85)$ where the $S$ and $I$ can take any values.  The initial state is $(S,I,R) = (95,5,0)$. 

We performed $10^6$ samples using an Intel Core i7-2620M CPU at 2.7GHz and computed the running time in seconds for various values of $\epsilon$, the results of which are shown in Table 1.  As can be seen, the running time is reduced when $\epsilon$ takes higher values since this will effectively reduce the number of gamma distributed samples needed to compute the exit time.  As $\epsilon \rightarrow 0$, the method reduces to the classical stochastic simulation algorithm.  

\begin{table}
  \begin{tabular}{|c| c | c | c | c| }
\hline
   $\epsilon$ & $ 0.5$ & $ 0.25$ & $0.125$  & $0$ \\ \hline
   Running Time [s] & 2.465 & 2.502 & 2.521 & 3.03\\ 
    \hline
  \end{tabular}
\caption[Table caption text]{Total simulation running time shown in seconds, $\epsilon=0$ denotes the standard simulation algorithm.}
\end{table}

\begin{figure}[htp]
  \begin{center}
     \subfigure{\includegraphics[width= 0.4\textwidth] {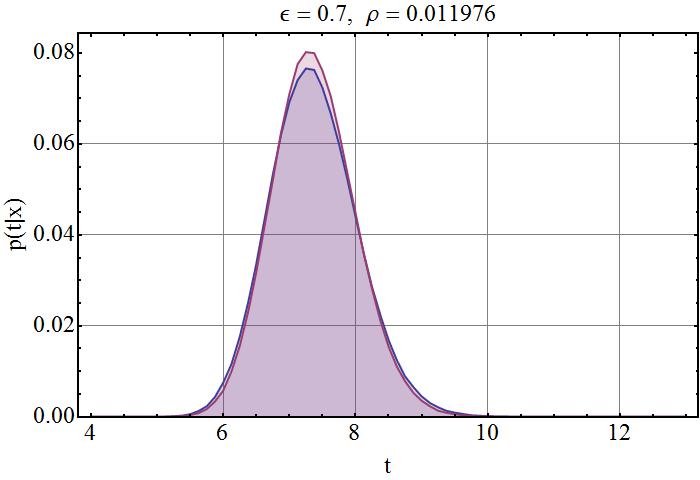}} 
    \subfigure{\includegraphics[width= 0.4\textwidth] {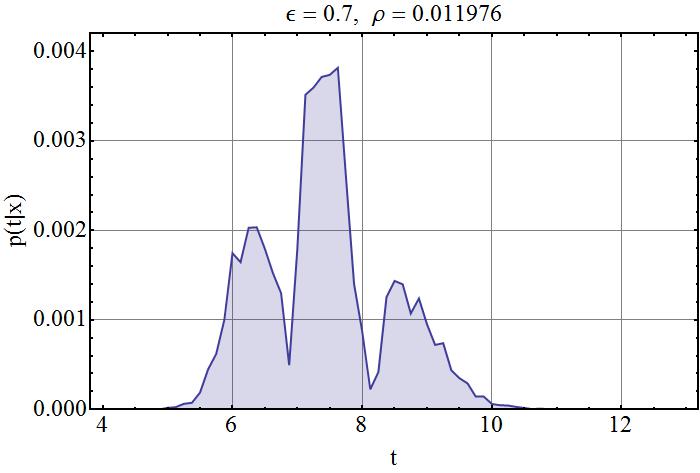}} 
  \subfigure{\includegraphics[width= 0.4\textwidth] {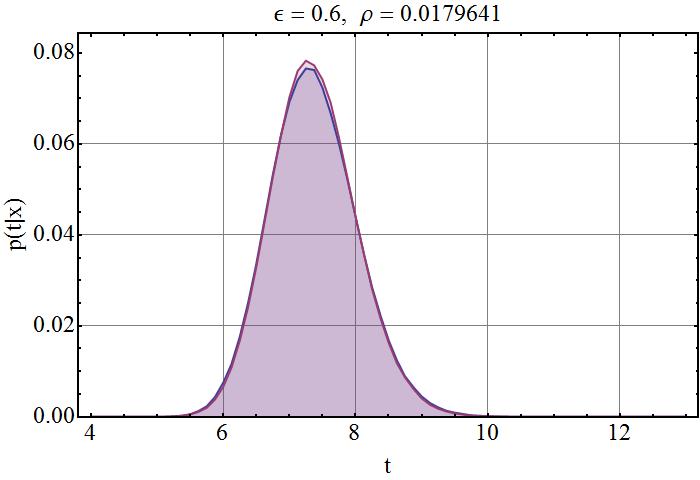}}  
     \subfigure{\includegraphics[width= 0.4\textwidth] {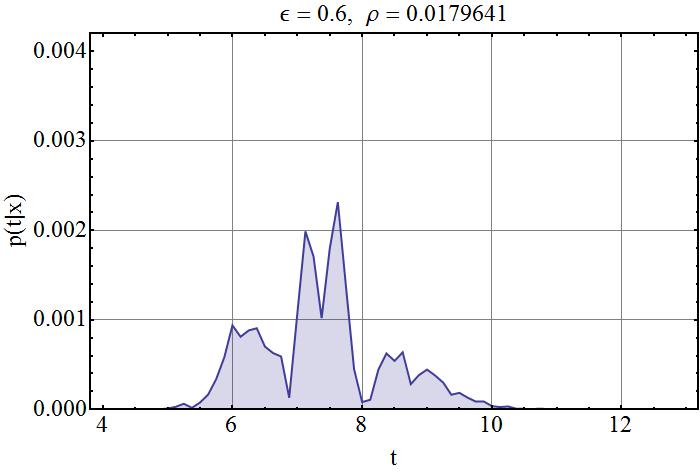}} 
  \subfigure{\includegraphics[width= 0.4\textwidth] {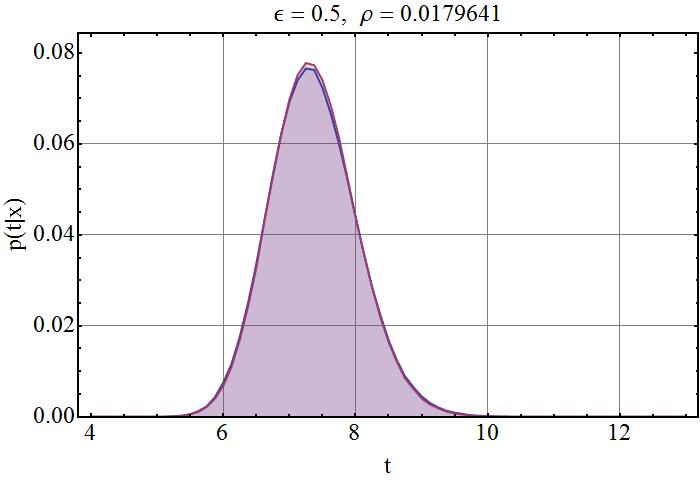}}  
     \subfigure{\includegraphics[width= 0.4\textwidth] {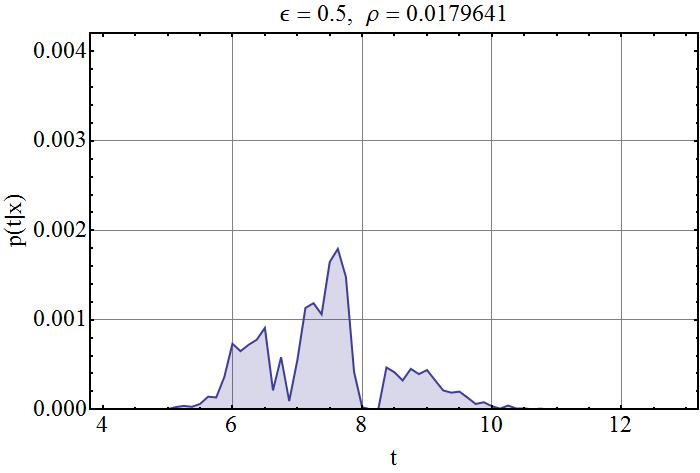}} 
  \subfigure{\includegraphics[width= 0.4\textwidth] {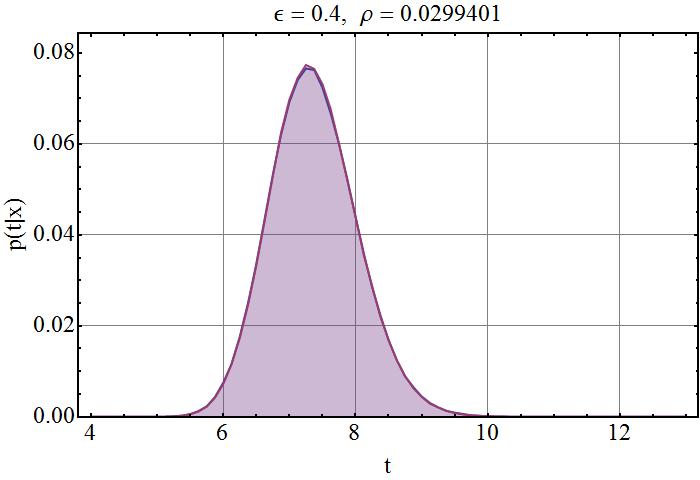}}  
     \subfigure{\includegraphics[width= 0.4\textwidth] {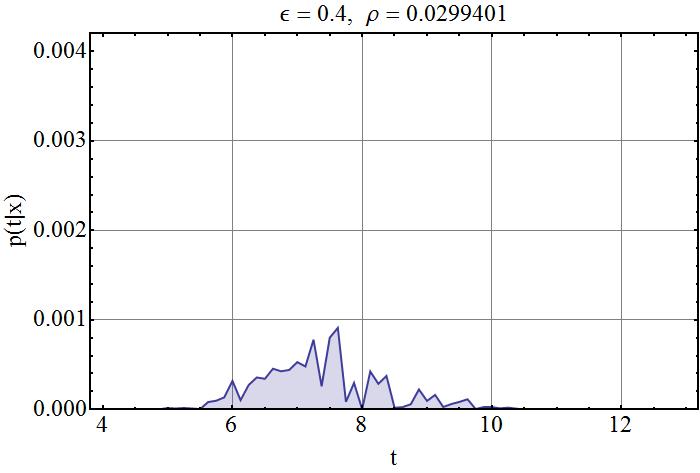}} 
     	\caption[]{Density function (left column) and point-wise error (right column) compared to $10^6$ SSA trajectories.  In the left column, the method presented above is shown in light red, while SSA is shown in purple.  The top figures denote larger values of the error control parameter $\epsilon$, while the lower figures show smaller values.  $\rho$ denotes the ratio of gamma distributed random variables to exponential random variables.  Note that the Monte Carlo sampling error is $\mathcal{O}\left( 10^{-3} \right)$.    }
	 \label{fig:PDFerror} 
  \end{center}
 \end{figure}

\begin{figure}[htp]
  \begin{center}
     \includegraphics[width= 0.75\textwidth] {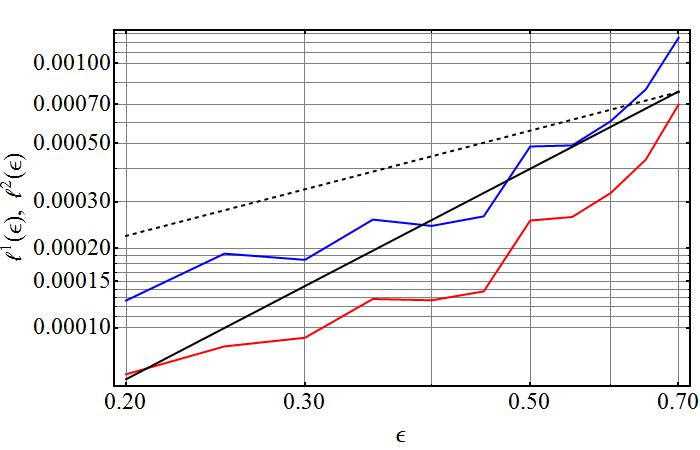}
     	\caption[]{Error of the exit time method: $l^1$ norm in red, and $l^2$ norm in blue, $\mathcal{O}(\epsilon)$ dashed and $\mathcal{O}\left( \epsilon^2 \right)$ solid lines.  Note that the Monte Carlo error is $\mathcal{O}\left( 10^{-3} \right)$. }
	 \label{fig:convergence} 
  \end{center}
 \end{figure}

\begin{figure}[htp]
  \begin{center}
     \includegraphics[width= 0.75\textwidth] {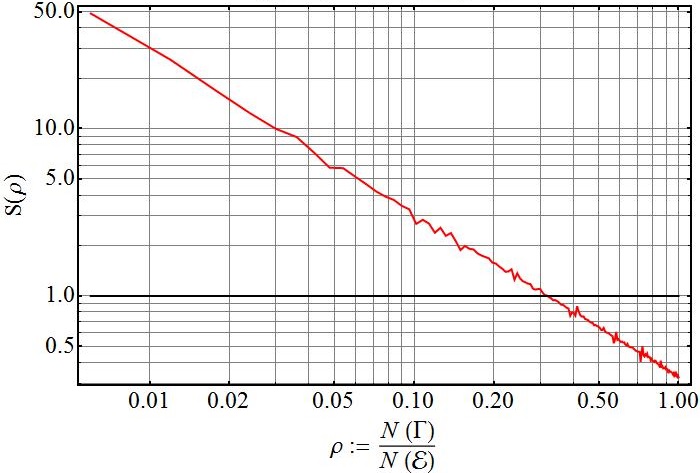}
     	\caption[]{Random variable generation speedup: Speedup of gamma distributed samples compared to exponentially distributed.  In the SIR example, $\rho$ ranges from $0.01 - 0.03$, yielding a speedup of $S(\rho)$ is between $10-30$.  Note that this is the speedup of the random number generation not the total simulation time.}
	 \label{fig:speedup} 
  \end{center}
 \end{figure} 

Since increasing $\epsilon$ will increase the error, we compared the density function of the exit times from the method presented here and the classical simulation algorithm in Figure \ref{fig:PDFerror}.  We have shown the two distributions in the left panel as well as the difference between them on the right.  We note that the exit time method is able to capture the correct distribution.  In addition, we reported the value for $\rho \triangleq N(\Gamma)/N(E)$ which is the number of gamma distributed random variables needed for the exit time method divided by the number of exponential random variables needed for the stochastic simulation algorithm.  In Figure \ref{fig:convergence} we have plotted convergence with respect to $\epsilon$, which is in accord with the error analysis included in the derivation in section 2.  We note that there is a tapering off in the error which is incident to the Monte Carlo error becoming larger.  In Figure \ref{fig:speedup} we have shown the speedup where $\rho$ is the ratio of the number of gamma random variates to the number of exponential random variates.  We note that this is not the speedup of the simulation, but merely for the computation of random numbers.

\section{Conclusion}

A original method was presented to compute the exit time for the stochastic simulation algorithm.  The method was based on the addition of a series of random variables and was derived using the convolution theorem.  We derived the final distribution and showed one approximation method.  The result led to a formulation of the stochastic simulation algorithm that requires fewer random variates.   As shown in the results section above for a typical nonlinear model, the error control parameter $\epsilon$ can be suitably chosen such that the number of random variates needed to resolve the exit time is reduced.  Equation \eqref{eq:finalSample} has better convergence properties than leaping algorithms since the error is second order whereas leaping algorithms are typically first order, therefore, the Monte Carlo error will still dominate the total error.  While the method is similar to R-Leaping \cite{Auger:2006} in that the distribution for time is drawn from a Gamma distribution, it differs in that the propensities are grouped according to their magnitude and the trajectory for the state transitions is exact.  

Although the number of random variates has been reduced, the bottleneck of the stochastic simulation algorithm is still the re-computation of the propensities at each time-step.  While we report relatively modest speed-up compared to the classical stochastic simulation algorithm, the derivation and application of the method may be beneficial to other areas of algorithmic research.  Indeed the derivation is not limited to the stochastic simulation algorithm and in principle could be used in other Monte Carlo algorithms.   If $Y_{i}$ is a sequence of random variables for $i = 1, \ldots, n$ such that $Y_{i} \sim \mathcal{D}(\Theta_i)$, where $\Theta_i$ is independent of $Y_1, \ldots, Y_n$, then the derivation could be used to find an expression for $Y_1 + \ldots + Y_n$.  

\section{Acknowledgements}
The author thanks Bill and Melinda Gates for their active support of this work and their sponsorship through the Global Good Fund.   

\appendix
\section{Appendix}
\label{sec:Appendix}
This appendix is provided for completeness and includes derivations of mathematical results used in the exit time method presented in section 2. 
\subsection{Hypoexponential Distribution}
\label{subsec:HypoexponentialDistribution}

\subsubsection*{Derivation of the Probability Density Function}
\label{subusb:DerivationOfHypoexponentialDistribution}

We begin by attempting to rewrite the product in terms of partial fractions: 
\begin{eqnarray}
 \prod_{i=1}^n  \frac{\lambda_i}{\lambda_i + s} \stackrel{\text{!}}{=} \sum_{i=1}^n \frac{C_i}{\lambda_i + s }
\end{eqnarray}
where $C_i \triangleq C_i(\lambda_1, \ldots, \lambda_n)$ must be determined and $ \stackrel{\text{!}}{=}$ denotes `shall be equal to'.  Then, 
\begin{eqnarray}
 \prod_{i=1}^n  \frac{\lambda_i}{\lambda_i + s} &=& \sum_{i=1}^n \left( \frac{C_i }{\left(\lambda_i + s\right)} \frac{ \prod_{\substack{j=1\\j \neq i}}^n \left( \lambda_j + s \right)}{ \prod_{\substack{j=1\\j \neq i}}^n \left( \lambda_j + s \right)}  \right) \implies \\
 \prod_{i=1}^n  \lambda_i &=&  \sum_{i=1}^n \left(C_i  \prod_{\substack{j=1\\j \neq i}}^n \left( \lambda_j + s \right) \right),
\end{eqnarray}
which holds $\forall s$.  If $\mathfrak{Re}(s) = -\lambda_1$ and $\mathfrak{Im}(s) = 0$, then 
\begin{eqnarray}
 \prod_{i=1}^n  \lambda_i &=& C_1  \prod_{\substack{j=1\\j \neq 1}}^n \left( \lambda_j  - \lambda_1 \right) +  \underbrace{C_2  \prod_{\substack{j=1\\j \neq 2}}^n \left( \lambda_j  - \lambda_1 \right)}_{= 0} +  \underbrace{\ldots}_{=0} \\
C_1 &=& \frac{ \prod_{i=1}^n  \lambda_i}{ \prod_{\substack{j=1\\j \neq 1}}^n \left( \lambda_j  - \lambda_1 \right) } \\
&=& \lambda_1 \prod_{\substack{j=1\\j \neq 1}}^n \frac{\lambda_j}{\lambda_j - \lambda_1}
\end{eqnarray}
therefore 
\begin{eqnarray}
C_i = \lambda_i \prod_{\substack{j=1\\j \neq i}}^n \frac{\lambda_j}{\lambda_j - \lambda_i}
\end{eqnarray}
\begin{eqnarray}
l_i \triangleq l_i\left(\lambda_1, \ldots, \lambda_n \right) \triangleq \prod_{\substack{j=1\\j \neq i}}^n \frac{\lambda_j}{\lambda_j - \lambda_i}
\end{eqnarray}
then 
\begin{eqnarray}
 \prod_{i=1}^n  \frac{\lambda_i}{\lambda_i + s} &=& \sum_{i=1}^n l_i \frac{\lambda_i}{\lambda_i + s}.  
\end{eqnarray}
In order to find the analytical form of the original convolution, we apply the inverse Laplace transform, viz.:
\begin{eqnarray}
\mathcal{L}^{-1}\left\{  \prod_{i=1}^n  \frac{\lambda_i}{\lambda_i + s} \right\}(t) &=& \mathcal{L}^{-1}\left\{ \sum_{i=1}^n l_i \frac{\lambda_i}{\lambda_i + s} \right\}(t) \\
&=&  \sum_{i=1}^n  l_i  \mathcal{L}^{-1}\left\{\frac{\lambda_i}{\lambda_i + s} \right\}(t) 
\end{eqnarray}
The inverse can be found by noting that $ \mathcal{L}^{-1}\left\{ \mathcal{L}\left\{ f(t) \right\}(s) \right\}(t) = f(t)$ and therefore $\mathcal{L}^{-1}\left\{\frac{\lambda_i}{\lambda_i + s} \right\}(t) = \lambda_i e^{-\lambda_i t}$.  

\subsubsection*{Summary (see \cite{Bolch:2006}):}
\begin{eqnarray}
\mathcal{L}^{-1}\left\{  \prod_{i=1}^n  \frac{\lambda_i}{\lambda_i + s} \right\}(t) =  \sum_{i=1}^n  \left(  \prod_{\substack{j=1\\j \neq i}}^n \frac{\lambda_j}{\lambda_j - \lambda_i} \right)   \lambda_i e^{-\lambda_i t},
\end{eqnarray}

\subsubsection*{Drawing Random Variates from the Probability Density Function}
We use the inversion theorem \cite{Devroye:1986} to obtain: 
\label{sec:RandomVariatesFromPDF}
\begin{eqnarray}
p(t; \lambda_1, \ldots, \lambda_n) &=& \sum_{i=1}^n  l_i  \lambda_i e^{-\lambda_i t} \\
\int_0^{t} p(t'; \lambda_1, \ldots, \lambda_n)~\text{d}t' &=& \int_0^{t} \sum_{i=1}^n  l_i  \lambda_i e^{-\lambda_i t'}  ~\text{d}t' \\
&=& \sum_{i=1}^n  l_i  \int_0^{t}  \lambda_i e^{-\lambda_i t'}  ~\text{d}t' \\
&=& \sum_{i=1}^n  l_i  \left( 1-e^{-\lambda_i t} \right) 
\end{eqnarray}
Define the cumulative distribution $\hat{p}(t; \lambda_1, \ldots, \lambda_n) =  \sum_{i=1}^n  l_i  \left( 1-e^{-\lambda_i t} \right)$, then, find $t$ such that $ \sum_{i=1}^n  l_i  \left(1-e^{-\lambda_i t}\right)  = r$, where $r \sim \mathcal{U}(0,1)$ by a root finding method.

\subsection{Frequency-Domain of Gamma/Erlang Distribution}
\label{sec:GammaFrequency}
Define the gamma distribution: 
\begin{eqnarray}
q(t; \lambda, n) \triangleq \frac{\lambda^n t^{n-1} e^{-\lambda t}}{\Gamma(n)},  
\end{eqnarray}
where $t \in [0, \infty)$, $\lambda > 0$, $n >0$, and $\Gamma(n) = (n-1)!$. Laplace's transform yields (`i.b.p.' denotes integration by parts):
\begin{eqnarray}
\mathcal{L}\left\{ q(t; \lambda, n) \right\}(s) &=& \int_0^{\infty} \frac{\lambda^n t^{n-1} e^{-\lambda t}}{\Gamma(n)} e^{-s t}~\text{d}t \\
&=& \frac{\lambda^n}{\Gamma(n)} \int_0^{\infty} t^{n-1} e^{-t (\lambda+s)}~\text{d}t \\
&\stackrel{\text{i.b.p}}{=}& \frac{\lambda^n}{\Gamma(n)} \int_0^{\infty} (n-1) t^{n-2} \frac{1}{(\lambda + s)} e^{-t (\lambda+s)}~\text{d}t \\
&=& \frac{\lambda^n}{(\lambda + s)}\frac{1}{\Gamma(n)} (n-1)  \int_0^{\infty} t^{n-2} e^{-t (\lambda+s)}~\text{d}t  \\
&\stackrel{\text{i.b.p}}{=}& \frac{\lambda^n}{(\lambda + s)^2}\frac{1}{\Gamma(n)}  (n-1) (n-2) \int_0^{\infty} t^{n-3} e^{-t (\lambda+s)}~\text{d}t~~~~ \\
&\stackrel{\text{i.b.p}}{=}& \ldots \\
&=& \frac{\lambda^n}{(\lambda + s)^n}\frac{1}{\Gamma(n)} (n-1)! \\
&=&  \left(  \frac{\lambda}{\lambda + s} \right)^n
\end{eqnarray}

\subsubsection*{Summary:}
\begin{eqnarray}
\mathcal{L}\left\{ \frac{\lambda^n t^{n-1} e^{-\lambda t}}{\Gamma(n)}    \right\}(s)  =  \left(  \frac{\lambda}{\lambda + s} \right)^n
\end{eqnarray}

\subsection{The Residue Theorem}
The residue theorem \cite{residueMathworld} can be used to compute the inverse Laplace transform:
\label{sec:residueTheory}
\begin{eqnarray}
f(t) = \mathcal{L}^{-1}\left\{ \tilde{f}(s) \right\}(t) = \sum_k \text{Res} \left( \tilde{f}(s) e^{s t}, s_k \right), 
\end{eqnarray}
where $s_k$ is a pole and the complex residue for a pole $s_k$ of order $\eta$ is 
\begin{eqnarray}
 \text{Res} \left( \tilde{f}(s) e^{s t}, s_k \right) = \frac{1}{(\eta-1)!}  \lim_{s \rightarrow s_k} \frac{\text{d}^{\eta-1}}{\text{d}s^{\eta-1}} \left( (s-s_k)^{\eta}  \tilde{f}(s) e^{s t}  \right)
\end{eqnarray}

\subsection{Approximating the Density Function}
\label{sec:Approx}
Let   
\begin{eqnarray}
f(\xi, s) \triangleq \frac{\xi}{\xi + s}
\end{eqnarray}
and let $\lambda_1 \triangleq \lambda + \epsilon$ and $\lambda_2 \triangleq \lambda - \epsilon$, where $\epsilon \ll 1$.  Then, we can write 
\begin{eqnarray}
f(\lambda_1, s) f(\lambda_2, s) = f(\lambda, s)^2 + \mathcal{O}\left(\epsilon^2 \right).  
\end{eqnarray}
This can be seen by expanding $f(\lambda_1, s)$ and $f(\lambda_2, s)$: 
\begin{eqnarray}
f(\lambda_1, s) = f(\lambda + \epsilon, s) = f(\lambda, s) + \epsilon \frac{\partial f}{\partial \lambda} + \frac{\epsilon^2}{2} \frac{\partial^2 f}{\partial \lambda^2} + \mathcal{O}\left(\epsilon^3\right).  
\end{eqnarray}
and 
\begin{eqnarray}
f(\lambda_2, s) = f(\lambda - \epsilon, s) = f(\lambda, s) - \epsilon \frac{\partial f}{\partial \lambda} + \frac{\epsilon^2}{2} \frac{\partial^2 f}{\partial \lambda^2} - \mathcal{O}\left( \epsilon^3 \right).  
\end{eqnarray}
then 
\begin{eqnarray}
f(\lambda_1, s) f(\lambda_2, s) = \left( f(\lambda, s) + \epsilon \frac{\partial f}{\partial \lambda} + \frac{\epsilon^2}{2} \frac{\partial^2 f}{\partial \lambda^2} + \mathcal{O}\left(\epsilon^3 \right)  \right) \times \\ \left( f(\lambda, s) - \epsilon \frac{\partial f}{\partial \lambda} + \frac{\epsilon^2}{2} \frac{\partial^2 f}{\partial \lambda^2} - \mathcal{O}\left(\epsilon^3\right) \right)~~~\\
=  f(\lambda, s)^2 +  \epsilon \frac{\partial f}{\partial \lambda}  f(\lambda, s) +  \frac{\epsilon^2}{2} \frac{\partial^2 f}{\partial \lambda^2}   f(\lambda, s)   -   \epsilon \frac{\partial f}{\partial \lambda}  f(\lambda, s) \\ -  \epsilon^2 \left(  \frac{\partial f}{\partial \lambda} \right)^2   \frac{\epsilon^2}{2} \frac{\partial^2 f}{\partial \lambda^2}f(\lambda, s) +  \mathcal{O}\left(\epsilon^3\right) \\
=  f(\lambda, s)^2 + \mathcal{O}\left(\epsilon^2\right)
\end{eqnarray}

\subsubsection*{Summary:}
\begin{eqnarray}
\left(\frac{\lambda_1}{\lambda_1 + s}\right) \left(\frac{\lambda_2}{\lambda_2 + s}\right) = \left( \frac{\lambda}{\lambda + s} \right)^2 + \mathcal{O}\left( \epsilon^2 \right), 
\end{eqnarray}
where  $\lambda_1 \triangleq \lambda + \epsilon$ and $\lambda_2 \triangleq \lambda - \epsilon$.  

\subsection{Determining the $\tilde{\lambda}_k$s and $n_k$s for a given $\epsilon$}
\label{sec:Epsilon}

We will partition the set  $\Lambda = \{\lambda_1, \ldots, \lambda_n \}$ into $m$ disjoint sets $A_1, \ldots, A_m$, i.e.  $\bigcup_{k=1}^m A_k = \Lambda$ and $\bigcap_{k=1}^K A_k = \emptyset$.  

Initially set $A_k = \emptyset$ $\forall k$ and let $\hat{\Lambda}$ be the set $\Lambda$ sorted in descending order.  The procedure is as follows:  for $k = 1, \ldots, m$, set $A_k = \{ a | a \in \hat{\Lambda}~\wedge~a \ge  \hat{\Lambda}_1 - \epsilon \hat{\Lambda}_1  \}$, where $\hat{\Lambda}_1$ is the first element in the set $\hat{\Lambda}$, and then set $\hat{\Lambda} =\hat{\Lambda} - \bigcup_{j=1}^{k} A_j$ and repeat until $\hat{\Lambda} = \emptyset$.   Note that $m$ need not be known a priori and that $\epsilon = 0$ is valid.  

For each set $A_k$, calculate the mean rate $\tilde{\lambda}_k$:
\begin{eqnarray}
\frac{1}{\tilde{\lambda}_k} = \frac{1}{|A_k|}\sum_{i \in A_k} \frac{1}{\lambda_i}
\end{eqnarray}
and set $n_k = |A_k|$, which is to be used in equation \ref{eq:finalSample}, where $|.|$ denotes the number of elements of the set.

%
%


\end{document}